\documentclass[aps,prb,float,a4paper,twocolumn,superscriptaddress]{revtex4-1}
\usepackage{amsmath}
\usepackage{color}
\usepackage{floatrow}
\usepackage[rightcaption]{sidecap}
\usepackage{amssymb}
\usepackage{graphicx}

\begin{document}

\title{Twinning-induced pseudoelastic behavior in (MoW)$_{85}$(TaTi)$_{7.5}$Zr$_{7.5}$} 

\author{Aayush~Sharma}
\affiliation{Department of Mechanical Engineering, Iowa State University, Ames, IA 50011 USA}
\author{Valery~I.~Levitas}
\affiliation{Department of Mechanical Engineering, Iowa State University, Ames, IA 50011 USA}
\affiliation{Ames Laboratory, U.S. Department of Energy, Ames, IA 50011 USA}
\affiliation{Department of Materials Science \& Engineering, Iowa State University, Ames, IA 50011 USA}
\author{Prashant~Singh}
\affiliation{Ames Laboratory, U.S. Department of Energy, Ames, IA 50011 USA}
\author{Anup~Basak}
\affiliation{Department of Aerospace Engineering, Iowa State University, Ames, IA 50011 USA}
\author{Ganesh~Balasubramanian}
\affiliation{Department of Mechanical Engineering \& Mechanics, Lehigh University, Bethlehem, PA 18015 USA}
\author{Duane~D.~Johnson}
\affiliation{Ames Laboratory, U.S. Department of Energy, Ames, IA 50011 USA}
\affiliation{Department of Materials Science \& Engineering, Iowa State University, Ames, IA 50011 USA}

\begin{abstract} 
We provide a critical atomistic evidence of pseudoelastic behavior in complex solid-solution BCC Mo-W-Ta-Ti-Zr alloy. Prior to this work, only limited single-crystal BCC solids of pure metals and quaternary alloys have shown pseudoelastic behavior at low temperatures and high strain rates. The deformation mechanisms investigated using classical molecular simulations under tensile-compressive loading reveal temperature-dependent pseudoelastic behavior aided by twinning during the loading-unloading cycle. The pseudoelasticity is found to be independent of loading directions with identical cyclic deformation characteristics during uniaxial loading. Additionally, temperature variation from 77 to 1500 K enhances the elastic strain recovery in the alloy. \\
\textbf{Keywords:} deformation, pseudoelasticity, high entropy alloy, twinning, molecular simulations
\end{abstract}

\maketitle

Martensitic transformations (MTs) are first-order diffusionless transformations that are observed in ferroelectric and ferromagnetic alloys and play the central role in exhibiting the pseudoelasticity or superelasticity in shape-memory alloys.\cite{Kaushik2003,1_} Since the pioneering work of Kurdyumov and coworkers about the nature of the MTs,\cite{4,5} a deep understanding exists that has made it possible to create a new class of materials showcasing pseudoelasticity.\cite{6} Under mechanical load and reduction in temperature, the austenite (parent phase) transforms into a martensite (product phase), which consists of complex microstructures such as twinned martensite, wedges, and twins within twins.\cite{10,2_} When the load is removed and the temperature is increased, the martensite transforms back into the parent phase and the initial shape is recovered. Such special transformations are responsible for the pseudoelasticity in alloys.\cite{11} However, pinning of the austenite-martensite and martensite-martensite interfaces by dislocations or other crystal defects often gives rise to irreversible martensitic transformations that may inhibit complete pseudoelasticity.\cite{10,11}

Recently, much attention has been on identifying pseudoelastic alloys, due to their diverse application, especially as actuation devices and bio-compatible stents.\cite{Zhang2005} Eliminating the toxicity issue in some of existing pseudoelastic materials, e.g., Ni-Ti, poses a challenge to the biomedical industry.\cite{Zhou2004-01,Takahashi2002} All that being said, the complex microstructures of these pseudoelastic alloys make it difficult to experimentally characterize the reversible stress-induced deformation products. To discover an alternate non-toxic alloy, we perform extensive research on novel refractory-based Mo-W-Ta-Ti-Zr high entropy alloys (HEAs).\cite{Yeh2004} HEAs are of intense interest due to their remarkable mechanical behavior, structural strength, resistance to fatigue, oxidation, corrosion, and wear,\cite{Gao2016,Miracle2017,Miracle2017_1} with a potential view of employing these materials in systems ranging across defense equipment, naval architecture, and high-temperature applications.\cite{Gao2016, Singh2018}

Our recent investigation on novel refractory Mo-W-Ta-Ti-Zr complex solid-solution alloy (CSA) reveals interesting electronic and mechanical characteristics, alongside global and local stability.\cite{Singh2018,PRM} While sweeping through 5-dimensional composition space, we zeroed on one such composition, (MoW)$_{85}$(TaTi)$_{7.5}$Zr$_{7.5}$ (MWTTZ), which exhibits pseudoelastic deformation under applied strain. The understanding of the control mechanism in MWTTZ alloy, along with revealing the twinning process, makes an interesting case study.

\begin{figure}[t!]
\centering
\includegraphics[scale=0.25]{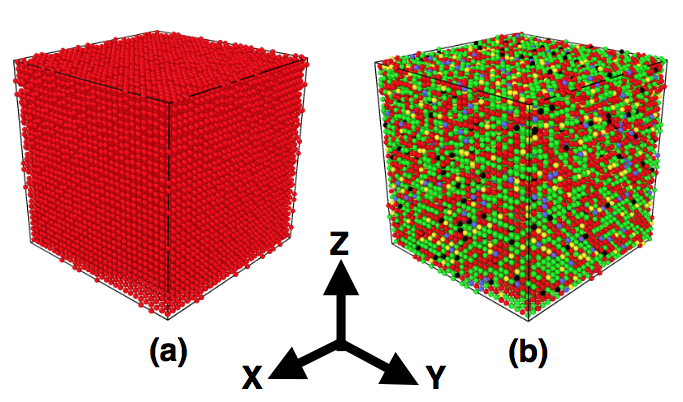}
\caption{Simulation cell of (a) Tungsten (W) and (b) quinary (MoW)$_{85}$(TaTi)$_{7.5}$Zr$_{7.5}$ (MWTTZ) alloy. Common neighbor analysis reveals a BCC coordination among the 54,000 atoms for both. Elements are visualized through different colors: Mo=green, W=red, Ta=yellow, Ti=black, Zr=blue.}
\label{struct}
\end{figure}

A cuboidal simulation domain is constructed by random distribution of Mo, Ta, W, Ti and Zr atoms in a BCC lattice (Fig.~\ref{struct}) of (95.9014 $\times$ 95.9014 $\times$ 95.9014) \AA$^{3}$, with a lattice constant of, a$_{0}$ = 3.19 \AA.\cite{Singh2018} The mole fractions of the different elements constitutes a total of 54000 atoms. Periodic boundary conditions are imposed in all the directions. The intermolecular interactions are described using the assimilated Embedded Atomic Method (EAM),\cite{Zhou2004, AS2016} and validated previously.\cite{Singh2018} We employ the Large-scale Atomic/Molecular Massively Parallel Simulator (LAMMPS) simulation package,\cite{Plimton1995} for calculations. For visualization, the common neighbor analysis (CNA)\cite{Stukowski2012} is used in ``OVITO" the Open Visualization Tool.\cite{Stukowski2010}

With large-scale atomistic simulations, we reveal twinning and detwinning phenomena under loading and unloading of the MWTTZ alloy. Similar loading-unloading hysteresis originating from twinning process has been found in nanocrystalline tungsten (W).\cite{Wang2015} In these materials, twin boundaries act as effective barriers to dislocation slip, which in turn increases the yield strength and ductility. Previously such behavior were limited to a couple of refractory-based quaternary alloys.\cite{Hagihara2016,Hagihara2017} However, we did not find any reports on pseudoelastic behavior in quinary (five-component) alloys. The deformation mechanisms for the quaternary (Ti-Nb-Ta-Zr) are found to be dependent on composition and vary between slip and twinning modes.\cite{Hagihara2016,Hagihara2017} MWTTZ is the first quinary CSA that exhibits desired characteristics of pseudoelastic deformation, and holds great promise for biomedical applications.

The energy minimization of the structure carried out using the conjugate-gradient algorithm with energy tolerance of 10$^{-15}$ and force tolerance of 10$^{-15}$ (eV/\AA) results in a geometrically optimized configuration for the HEA. The structure is initialized at 4000 K under an isothermal-isobaric (NPT) ensemble at a pressure of 0 MPa for 90 picoseconds (ps). The alloy is rapidly quenched from 4000 K to different temperatures (300 to 1500 K) in 10 ns, and the structure is equilibrated for an additional 90 ps. The quenched HEA is further simulated under the NPT and NVT (canonical) ensembles, successively for 10 ps and 20 ps, respectively. The pressure (0 MPa), and temperature constraints are imposed by the N\'{o}se-Hoover barostat and thermostat with coupling times for both set at 1 ps. Finally, the structure is equilibrated for 10 ps under the microcanonical ensemble (NVE) to complete the quenching process. A time step of 0.001 ps is maintained throughout all our simulations. A quasistatic uniaxial loading-unloading in the $<$100$>$ direction is applied to analyze the deformation mechanisms. At each loading step, we expand the simulation box at a rate of 0.01 ps$^{-1}$, and the strain expressed on the sample is the true strain. Subsequently, the deformed alloy is equilibrated under NPT and NVE ensembles for 90 and 50 ps, respectively, after each loading step.

For any alloy, analyzing the phase stability is an important criteria. Based on valence electron composition ($4 < VEC < 6$), size-effect ($\le$6.6\%), and mixing (formation) energy $\Delta E_f$ (-15 mRy $\leq$ $\Delta E_f$ $\leq$ 5 mRy) calculated from first-principles shows that the HEA is energetically stable, and will form a complex solid-solution.\cite{Singh2018} Throughout our calculations, we consider the solid-solution phase. 

In Fig. \ref{struct}, the crystal lattice of (a) pure tungsten (W), and (b) quinary MWTTZ is shown. The common neighbor analysis (CNA) performed on the undeformed MWTTZ alloy reveals a body centered cubic (BCC) coordination. 

\begin{figure*}[t]
\floatbox[{\capbeside\thisfloatsetup{capbesideposition={right},capbesidewidth=5.2cm}}]{figure}[\FBwidth]
{\caption{(a) Hysteresis found in (MoW)$_{85}$(TaTi)$_{7.5}$Zr$_{7.5}$ (MWTTZ) during quasistatic loading/unloading at 300K; (b) stepwise atomic-scale picture of twinning (loading) and detwinning (unloading); and (c) shows the first set of twins, the cut of (111) clearly shows the twin-plane while the twin-direction is $\langle112\rangle$.
In (a), MWTTZ shows the twin formation at 0.065 strain, which on further strain hardening gives rise to cross-twins (T19). Compression (unloading cycle), associated with rise in the stress, leads to detwinning (C1 to C11) .}\label{fig:tension-compression}}
{\includegraphics[scale=0.32]{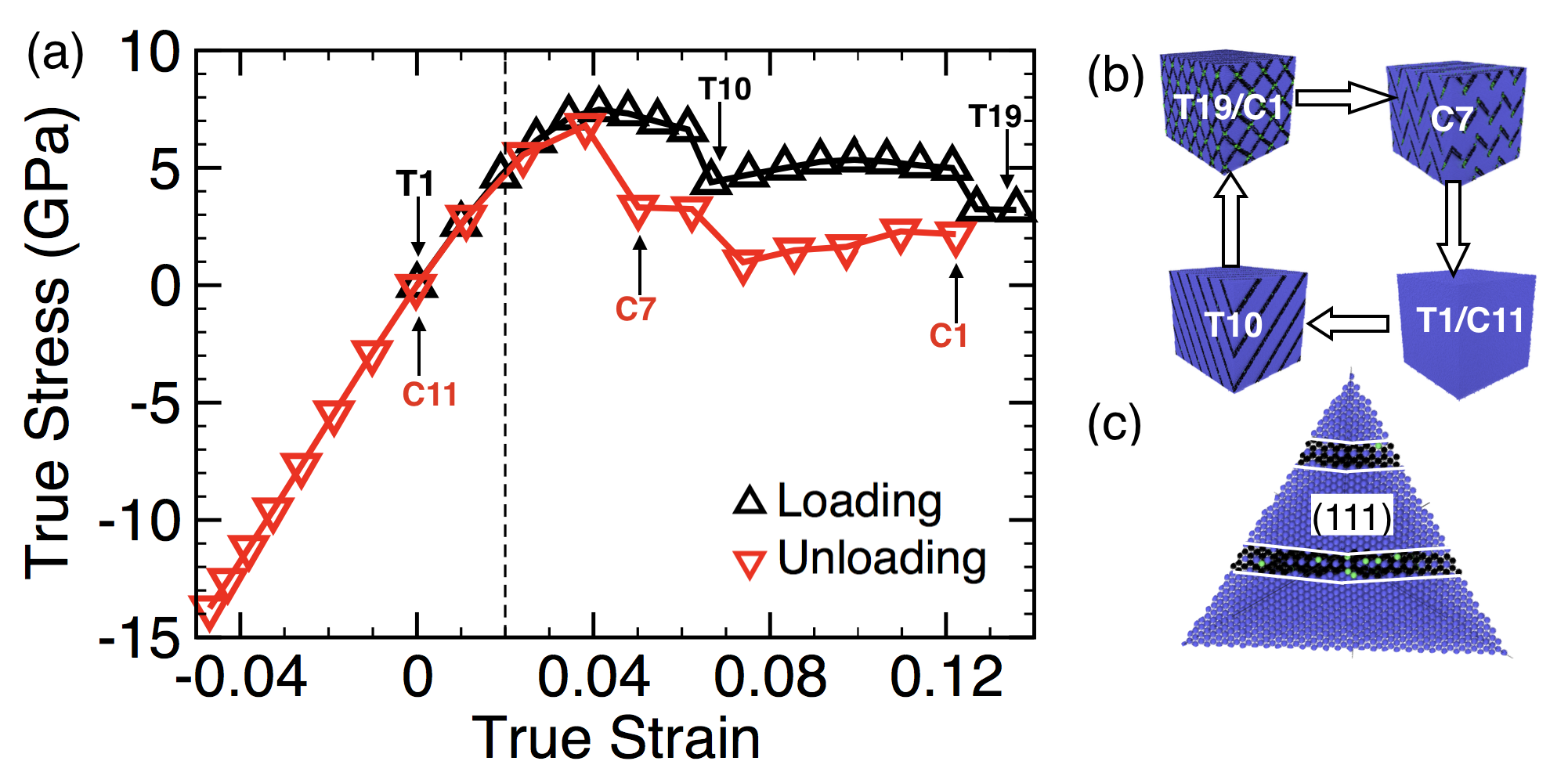}}
\end{figure*}

Alongside phase stability we also investigate the alloy's (MWTT) mechanical stability, based on standard criteria:\cite{Slaughter2002} 

\begin{equation}
B > 0; \quad {C}' > 0; \quad C_{44} > 0; \\
\label{Eqn:stability-1}
\end{equation}
where the bulk modulus ($B$) and shear modulus $C'$ are given by:
\begin{equation}
B = \frac{C_{11}+2C_{12}}{3}; {  \ \  \ \ }    {C}'=\frac{C_{11}-C_{12}}{2} \\
\label{Eqn:stability-2}
\end{equation}
C$_{11}$, C$_{12}$, and C$_{44}$ are three independent elastic constants in cubic crystals which relate all six components of the stress ('$\sigma$') tensor with six components of the strain ('$\epsilon$') tensor.\cite{Slaughter2002} These constants are usually derived from the total-energy calculations representing the single-crystal elastic properties. We find that the MWTTZ alloy satisfies the criteria for both (a) phase and (b) mechanical stability. 

\begin{figure}[b]
\centering
\includegraphics[scale=0.27]{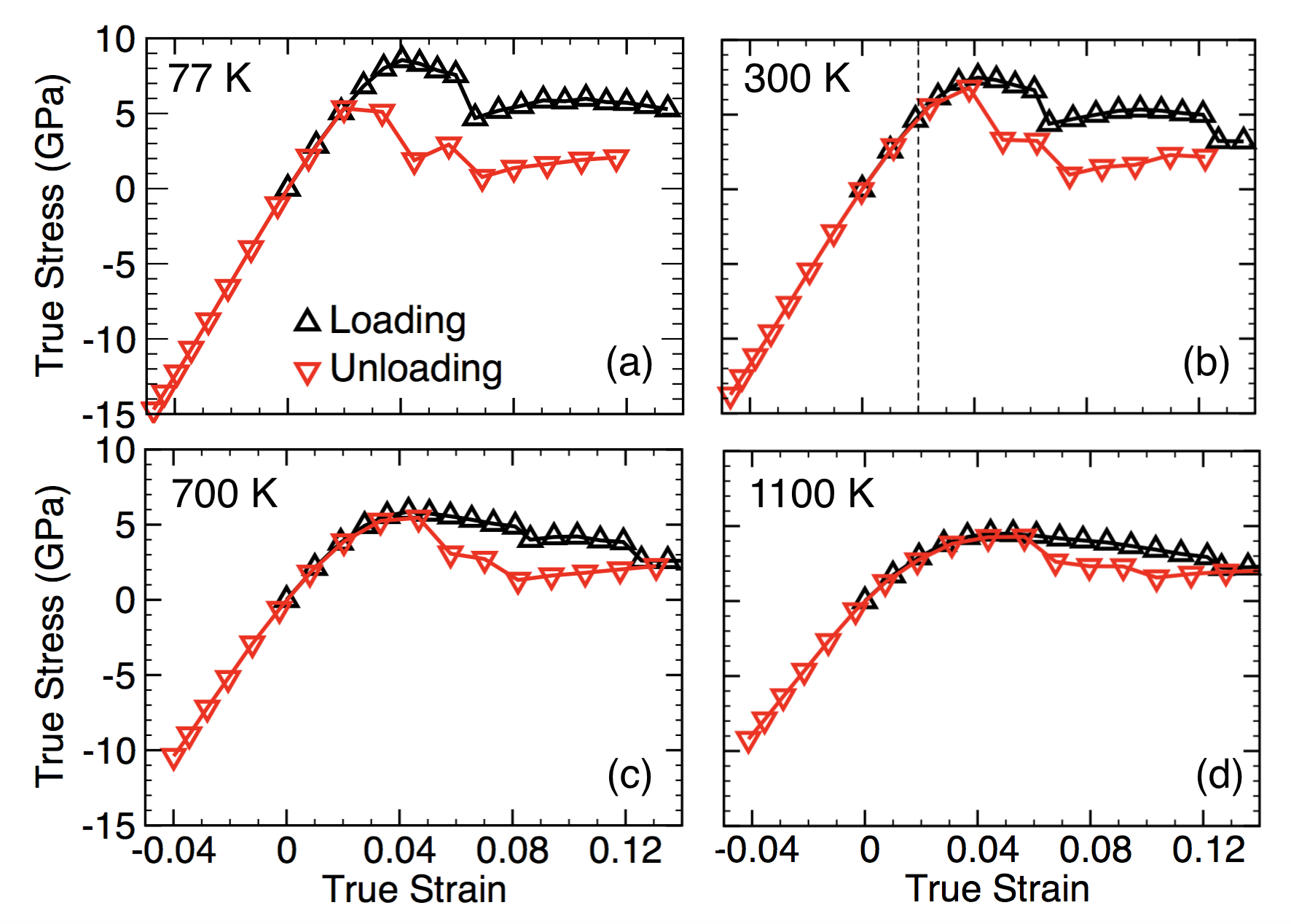}
\caption{Temperature-dependent pseudoelastic behavior observed in (MoW)$_{0.85}$(Zr(TaTi))$_{0.15}$ (MWTTZ) at (a) 77 K, (b) 300 K, (c) 700 K, and (d) 1100 K. The MWTTZ shows extreme thermal sensitivity. A large hysteresis found at 77 K slowly decreases with increase in temperature, and disappears at very high temperatures ($>$1100 K).}
\label{fig:thermal-stress-strain}
\end{figure}

To capture the correct values of the elastic constants and relevant parameters of MWTTZ, we consider the small-displacement method.\cite{Shinoda2004} We found a drastic drop in C$_{11}$ from 395 GPa at 77 K to 336 GPa at 1100 K, however, C$_{12}$ and C$_{44}$ show slight increase from 179$\rightarrow$191 GPa and 123 $\rightarrow$130 GPa, respectively, with increase in temperature. In a physical sense, C$_{11}$ shows reduction in longitudinal elastic behavior, whereas C$_{12}$ and C$_{44}$ show slow increase with temperature in off-diagonal and elastic shear characteristic of MWTTZ, respectively. A longitudinal strain produces a change in volume without a change in shape and is related to the pressure, which reflects a larger change in C$_{11}$. In contrast, a transverse strain or shearing causes a change in shape without a change in volume. So, C$_{12}$ and C$_{44}$ are less sensitive to pressure than C$_{11}$. The shear modulus (Eq.~\ref{Eqn:stability-2}) drops from 108 GPa at 77 K to 60.5 GPa at 1100 K, which clearly comes from the reduction in longitudinal elastic constant C$_{11}$.

We perform quasistatic tensile loading and unloading (compression) on the MWTTZ alloy to investigate the behavior of the alloy under external strain. Uniaxial loading is widely used to reveal the deformation characteristics of a material. We present one such deformation curve in Fig.~\ref{fig:tension-compression} at 300 K. MWTTZ follows elastic limit till 0.02 strain value with stress of 6.16 GPa. Beyond which, instability drives MWTTZ away from the elastic regime. As stress piles up, twinning is observed in the crystalline lattice (T10 in Fig.\ref{fig:tension-compression}). The first major stress drop is observed at T10 showing deviation from the perfect BCC coordination. At T10, with a major stress drop, twinning relieves the stress. The findings are also observed through the CNA mapping. The CNA analysis in the present study helps us to track the nucleation of twins, its growth, as well as detwinning in a loading-unloading cycle. Along with a first set of twins, we also observe a set of cross-twins in the MWTTZ alloy (T19 inset Fig. \ref{fig:tension-compression}). During unloading (compression) we find that the twins disappears (C7 inset Fig. \ref{fig:tension-compression}). Detwinning is characterized by an abrupt rise in stress levels and beyond C7, (C8 to C11 inset Fig. \ref{fig:tension-compression}), we find that the material is following its elastic loading curve.

The maximum shear prior to twinning (T9) was 0.17, while during twinning (T10), and for the second twin (cross-twin) it was 0.29, and 0.31 respectively. It is interesting to note from literature that BCC solids (1/2 atoms shuffle) show a twinning shear value of 0.35.\cite{Mahajan1995} 

To analyze the effect of thermal fluctuation on the pseudoelasticity observed in MWTTZ complex solid-solution alloy, we perform temperature (77 K to 1500 K) dependent loading-unloading deformations. In Fig.~\ref{fig:thermal-stress-strain}(a)\&(b), we show that the first twin formation is observed at $\approx$0.065 strain, while the cross-twins appear at $\approx$0.12 strain levels. With further increase in temperature, Fig.~\ref{fig:thermal-stress-strain}(c)\&(d), MWTTZ allows only cross-twins at higher strain ($\approx$0.12), which softens the elastic modes compared to low-temperature cases.\cite{Mahajan1973} Clearly, the large thermal fluctuations at higher temperature require higher external strain for the twin nucleation. The soften elastic modes at higher temperature, as shown in Fig.~\ref{fig:thermal-stress-strain}, leads to a reduction in the loading-unloading hysteresis. In MWTTZ, the temperature can be use as a control parameter to tune the pseudoelastic trend. Recent studies show that elemental `W' in single crystal form is known to exhibit deformation twins at negative temperatures (T $<$ 0$^\circ$C). Studies also suggest that an increase in the purity would facilitate the twinning process.\cite{Savitskii1970} In our MD analysis, we observe twinning during tensile loading for both pure tungsten, and the quinary MWTTZ {(see supplementary)}.

\begin{figure}[t]
\centering
\includegraphics[scale=0.4]{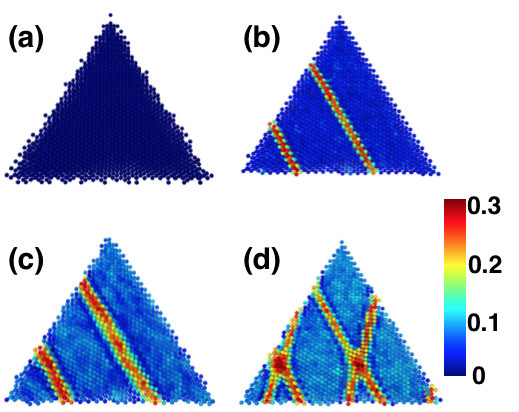}
\caption{Variation of shear strain during (tensile) loading cycle along $\left[111\right]$. (a) T1 (b) T9 (c) T17 and (d) T19 represents the quasistatic loading steps at different strains. Thickness of the twin layer grows from 3 atomic layers (T9) to around 4 atomic layers (T17). Further strain hardening leads to the formation of cross-twinning. The magnitude of stress required for initial twin nucleation is higher than that of its propagation.}
\label{fig:shear-strain}
\end{figure}

To investigate the evolution of shear strain for MWTTZ, we show twins along the (111) plane, in Fig.\ref{fig:shear-strain}. With no load, as anticipated, atoms experience zero shear (Fig.\ref{fig:shear-strain}a), with further increase in uniaxial loading ($<$100$>$ (x)) twins appear (Fig. \ref{fig:shear-strain}b; T9). The twinning layers are about 3 atomic layers thick which marginally increases to 4 atomic layers at T17 (Fig. \ref{fig:shear-strain}c). Any increase in quasistatic ($<$100$>$) load beyond this leads to the formation of additional cross-twins for the MWTTZ HEA. The thickness of both the original and the cross-twins is found to be approximately 3 atomic layers. The features of twins, cross-twins and reverse twinning or detwinning is reproduced for the uniaxial loading along any of the three directions: $<$100$>$ or $<$010$>$ or $<$001$>$. For clarity, we have specifically discussed the $<$100$>$ case. It is also worth mentioning that within the strain limits (small-strain regime) considered in the present study, biaxial loading ($<$110$>$ or $<$101$>$ or $<$011$>$) of the quinary MWTTZ alloy does not yield evidence of twinning and detwinning.


To conclude, we provide evidence for the pseudoelastic behavior in quinary high-entropy alloys using atomistic simulations for the very first time. Our calculations reveal the presence of pseduoelasticity in (MoW)$_{85}$(TaTi)$_{7.5}$Zr$_{7.5}$. We observe strong temperature dependent twinning and detwinning process during the loading-unloading (tension-compression) cycle. The twinning-detwinning feature is responsible for the pseduoelastic behavior in (MoW)$_{85}$(TaTi)$_{7.5}$Zr$_{7.5}$, which persists over a wide range of temperatures from 77 to 1500 K. In our findings, the elastic modulus softens with increasing temperature, which reduces the hysteresis and possibly can be used as a control parameter. Psuedoelastic materials have many possible applications including but not limited to actuators in devices and/or bio-medical implants. 

\acknowledgments
AS and GB thank the Office of Naval Research (ONR) for support through the grants N00014-16-1-2548 and N00014-18-1-2484.
The research was supported by a grant of computer time from the DoD High Performance Computing Modernization Program at Army Engineer Research and Development Center. The work at the Ames Laboratory was funded by the U.S. Department of Energy (DOE),  Office of Science, Basic Energy Sciences, 
Materials Science and Engineering Division. Ames Laboratory is operated for the U.S. DOE by Iowa State University 
under Contract No. DE-AC02-07CH11358.


\end{document}